\documentstyle[12pt,aaspp4]{article}
\begin{document}

\title{\bf QSO hosts and environments at z = 0.9 to 4.2: $JHK$ images 
with Adaptive Optics}

\author{J.B. Hutchings\altaffilmark{1}, David Crampton\altaffilmark{1}, 
S.L. Morris\altaffilmark{1}, and D. Durand} \affil{Dominion Astrophysical
Observatory,
National Research Council of Canada,\\ Victoria, B.C. V8X 4M6, Canada}

\author{E. Steinbring\altaffilmark{1}} \affil{Dept of Physics and Astronomy,
University of Victoria}

\altaffiltext{1} {Visiting Astronomers, Canada-France-Hawaii Telescope,
operated by the National Research Council of Canada, the Centre de la
Recherche Scientifique de France, and the University of Hawaii}

\begin{abstract}

  We have observed nine QSOs with redshifts 0.85 to 4.16 at near-IR
wavelengths with {\it Pueo}, the adaptive optics bonnette of the
Canada-France-Hawaii telescope.  Exposure times ranged from 1500 to
24000s (mostly near 7000s) in $J$, $H$, or $K$ bands, 
with pixels 0\farcs035 on the sky.
The FWHM of the co-added images at the location of the quasars are
typically 0\farcs16.  Including another QSO published previously, we
find associated QSO structure in at least eight of ten objects, including
the QSO at z = 4.16. The structures seen in all cases
include long faint features which appear to be tidal tails.
In four cases we have also resolved the QSO host galaxy, but find them to be
smooth and symmetrical: future PSF removal may expand this result. 
Including one object previously reported, of the 
nine objects with more extended structure,
five are radio-loud, and all but one of these appear to be in a dense
small group of compact galaxy companions. The radio-quiet objects do
not occupy the same dense environments, as seen in the NIR.  
In this small sample we do not
find any apparent trends of these properties with redshift, over the
range 0.8 $<$ z $<$ 2.4. The colors of the host galaxies and companions
are consistent with young stellar populations at the QSO redshift. Our
observations suggest that adaptive optic observations in the visible
region will exhibit luminous signatures of the substantial
star-formation activity that must be occurring.

\end{abstract}

\keywords{quasars: general, galaxies: interactions, instrumentation: adaptive
optics}

\section{Introduction}

QSO host galaxies are well investigated at redshifts up to $\sim$0.6,
with both ground-based and $HST$ images, both at visible and near infrared
wavelengths (e.g.Hutchings \& Neff 1992, 1997; McLeod \& Rieke 1995;
Bahcall et al.  1997 and references therein). Much has been written on
the general evidence for tidal events being the trigger for nuclear
activity, the hybrid morphology of the QSO host galaxies, associated
star-formation, and the different environments of companion galaxies
between radio-loud and radio-quiet QSOs.

At higher redshift, there have been ground-based data that resolve
QSO hosts at redshifts in the range 2.0 to 2.5. These, particularly
the radio-loud ones, appear to be very luminous galaxies, with very high
star-formation rates, and with no counterparts in the present day universe
(e.g. Heckman et al. 1991; Lehnert et al. 1992; Hutchings and Neff 1997b;
Hutchings 1998; Aretxaga et al. 1995,8).
There are also indications that QSOs of all kinds occur in small dense groups
of companion galaxies also with high star-formation rates (Hutchings 1995,8;
Campos et al 1998).  There is a clear similarity to 
radio galaxies at comparable redshifts; they also are very luminous and
are located in dense environments (e.g., Rottgering 1997, Pentericci et al 1998)

The high redshift QSOs observed so far lack the spatial detail and the
statistical samples that are available for the lower redshift QSOs, so
several key questions cannot yet be answered. Primarily these concern
the cosmic evolution of the QSO phenomenon. We do not know how the
nature of QSO activity evolves with redshift:  do the sites of QSO activity
evolve as different environments become favourable; does the environment
evolve at all; or does the environment not matter? 
We also do not know if the trigger of
nuclear activity changes with redshift  (as the host galaxies age or
as the interaction rate changes), or if the luminosity and visibility of
QSO activity evolves. Is star-formation a cause,
effect, or parallel process in QSO events? Do radio-loud and radio-quiet 
QSOs evolve differently?

We have begun a program to obtain high resolution and high
signal-to-noise images of QSOs in the redshift range 1 $<$ z $<$ 4. The
data reported here are from near infrared images obtained with {\it
Pueo}, the Adaptive Optics Bonnette of the CFHT (Rigaut et al. 1998).
We have previously reported similar observations for the z = 1.1
optically-selected QSO 1055+019 (Hutchings et al. 1998). In this paper
we add 9 more to the sample, and discuss the individual cases and the
trends we see in the sample so far. As noted in Hutchings et al 1998,
our observing list is restricted to QSOs with nearby bright guide stars
for the AO system. Thus, while we are investigating objects with redshifts
larger than $\sim$0.8, the sample available at any date of observation
is not large, and our principal criteria have been to attempt
to match radio-loud and radio-quiet objects, and to cover a wide range of
redshift.

\section{Observations and data}

The data reported here were obtained with {\it Pueo} in January and
March 1998. Table 1 gives the list of objects and Table 2 the journal
of observations. All observations used a nearby guide star, as listed
in the table, and corrected images at the location of the QSOs
generally have FWHM$\sim$ 0\farcs2 or better.
 The camera used is KIR, a 1024$\times$ 1024 pixel
Hg:Cd:Te detector described in the CFHT website. This detector has
lower noise and less image retention than the 256 pixel MONICA camera
used previously (i.e. fewer reads are required to remove the effects of
bright objects in the detector). 
Pixels are 0\farcs035, giving a field of 36\arcsec, so
that the guide star is usually in the science image as well as the QSO
and its local companions. The observations were 5 minute integrations
performed in a non-repeating dither pattern covering several arcsec.
Conditions were good for both the runs. The uncorrected seeing was in
the range 0\farcs5 to 0\farcs9 FWHM, and the transparency was good for
all observations reported. All observations were made in standard $J, H, K$
filters. Performance in the $K$ band is somewhat compromised by the
additional thermal background from the AO system, but unfortunately no
$K'$ or $K_S$ filter was available for our observations.
Near-infrared photometric standards (Casali
and Hawarden 1992) were observed, and images were also obtained of
crowded stellar fields (M91 and M15) to calibrate the variation of the
PSF with distance and brightness of the guide star.

Table 2 summarizes the data. The FWHM were measured at the QSO location
(these are, of course, somewhat worse than at the guide star). In the
best data the FWHM typically vary from 0\farcs13 at $K$ to 0\farcs15 at
$H$ and 0\farcs20 at $J$. There were substantial differences during and
between nights and, on average, the results were better during the
March run than the January run.

The data were processed using `standard' near infrared techniques, but
in several cases independently and differently by more than one author. 
We describe two approaches. First, sky
frames were obtained as medians of the unshifted dither
frames. In cases where the saturated guide star is in the image, these
were edited out before the median  was computed, as otherwise they
usually leave a faint imprint. In most cases we were able to use sky frames
from the entire period of observation of an object. However, during
nights when the sky brightness was very variable, running medians
of data taken closely in time were used.  After
sky-subtraction the data frames were shifted to superpose the QSOs
exactly, and combined with sigma clipping.
(The shifts were done with linear interpolation, after checking that
more elaborate shifts produced the same result.) Although flat field images
of dome lights were obtained, it was found empirically that flat field
division did little to improve the sky flatness or noise, presumably
as the sky frames constitute a flat field, and so they
were generally not used.

The small pixels and the bright sky make it hard to see
faint extended features such as tidal tails or the outer
parts of galaxies.  Further, no very sharp morphological features were
detected in any of the QSOs or their companions. Thus, to improve
detection of faint features, we further smoothed the sky by editing out
the QSO and companions and fitting a spline surface to the remaining
sky. This was subtracted from the image to remove sky brightness
variation on scales of several arcsec or more. This improved our
ability to investigate and measure faint tails and flux extensions over
scales up to a few arcsec. \it The above process was repeated
several times with slightly different edits and surface fits, and the
final result was a mean of the individual results, to remove spurious
artifacts. \rm  The features we report here are seen in
all processed images even before this flattening procedure, so we have
good confidence in their reality. Finally, in most cases we found that 
images with 4 $\times$ 4 block-averaged pixels (0\farcs14 pixels), were 
able to reveal the faint extended features more clearly.

   A similar procedure was used developed and used independently by
one of us, to deal with changing sky brightness and the faint haloes
around the bright guide stars. First, a median sky frame
was generated from all target exposures on each QSO for each filter. 
After correction
with this first-order flat-field, the exposures were grouped and combined
according to mean sky-brightness, and a mask was generated which included 
all bright pixels. Second-order flat-fields were then generated using the 
masks and applied for each group. Finally, the frames were registered
and median-combined.

   The features we discuss are faint and difficult to display, and in some
cases are comparable with sky features. However, we note that they are
seen independently of the different authors' processing, often appear in 
more than one filter, are robust to sky smoothing processing, and have 
flux measures that agree from differently processed images. Features that
did not pass all these criteria were rejected.

   Flux measurements were made by summing the signal in pixels that include the
structures of interest, and subtracting the mean nearby sky level. This was done
on both the full resolution and binned data. All galaxies and symmetrical
features were also measured by using the IRAF task imexam. Limiting
surface brightnesses vary with filter and integration time, as well as
sky conditions, but our rough surface brightness limits in mag per square
arcsec are 22.4 in $K$, 23.4 in $H$, and 24.4 in $J$.

We describe below the individual QSOs and their environments, and end
with some collective remarks on the ensemble.  The tables for each
object field give all measures made in this way.  Generally, the
magnitude errors for smooth round objects are $\pm$0.1 mag for $J$ =
19.0, $H$ = 18.4, $K$ = 17.9, and these increase to $\pm$0.2 mag for
objects 2.3 mag fainter and $\pm$0.4 mag for objects yet another 1 mag
fainter. Errors in extended features are comparable, while those for
PSF-subtracted fluxes are discussed individually. In this paper we have not
attempted to do careful PSF modelling and removal or deconvolution.
That will be dealt with in a separate paper. We mention the results of
simple PSF subtraction in a couple of cases where the QSO is sufficiently well
resolved that this is not a significant issue.

  We quote absolute magnitudes at the QSO redshift, for objects that have 
the size, brightness,
and colors to be possible companions. These are for an assumed H$_0 =
75$km s$^{-1}$ kpc $^{-1}$ and q$_0$ = 0.5, with no K-correction
(K-corrections depend on stellar population and bandpass, but for young
populations at these redshifts they are likely to be close to zero:
-1 mag for B0 stars and zero for A0 stars).
Comparison of these luminosities indicates that the possible companions
(and QSO host galaxies) have rest wavelength values in the range
typical of bright galaxies. If they have active star-formation, they
will be more luminous at shorter rest wavelengths. The shortest rest
wavelengths we are observing range from 4600\AA\ to 7000\AA, for the
three $J$ band observations. In $H$ band, these range from 4900\AA\ to
9000\AA.

   Figure~\ref{map} shows the distribution of detected objects in all the
fields. Measures of most of them are given in individual tables. The figure
indicates which objects are assumed from their size and/or brightness to be
foreground galaxies (F), and stars (*), as well as
the guide stars when they appeared in the field. We discuss them below in order
of increasing redshift.

\section{0915$-$213}

This is a radio-loud QSO at z = 0.85. 
It has a flat radio spectrum, of medium flux (0.6 Jy at 5 GHz), and is
probably compact, but there is no published radio map.
Our data are in the $H$ and $K$ bands. In both, the QSO lies 
in a group of four faint objects that occupy one quadrant of the KIR 
image - the rest of the field is empty, with the exception of two larger
low-surface-brightness galaxies, (and two stars) which are widely spaced. 
We suppose these to be foreground galaxies with low surface brightness.

The QSO is extended asymmetrically to the NE, and lies very close  (1\farcs4)
to the faintest compact companion (\#3), to the NW. A fainter companion 
(\#1) lies further to the N (see Figure~\ref{fig1}). 
There is an extended wisp of luminosity
beyond the line to the nearby companion, which turns through 90\arcdeg\ and
brightens slightly, some 4\arcsec\ from the QSO. These are very faint
features but they show up on subsets of our data and also in both colors,
so we regard them as real. Table 3 shows photometric measures of
objects in the field of 0915$-$213.

We have not yet obtained good PSF subtraction in the central QSO, either
because
of real complex structure, or because the PSF distortion away from the
guide star is unusually large. However, preliminary results indicate that the
QSO host is elliptical with its long axis in the direction of the
large extended luminosity to the NE. The fluxes given for the QSO host
are derived from the subtractions and also from the azimuthally averaged
profiles, which agreed to within 30\%.

The host galaxy extension differs between $H$ and $K$, suggesting a color
gradient. The nearby companion does not have a bright nucleus and is
extended tangentially to the direction to the QSO. While signal levels 
are low, we see no structure within the features described, and their
overall structure is very suggestive of tidal extension of an old
population of stars associated with the QSO host galaxy 
(see e.g. 2141+175 in Hutchings et al 1994) .

\section{0804+499}

This QSO, at z = 1.43, was observed in $H$ and $J$ bands. It is a radio
source that is unresolved at subarcsecond levels. The field is quite
empty, containing only the guide star, another object comparable with
the QSO, and a single faint galaxy some 20\arcsec\ away in both $H$ and
$J$  band.
In $J$ band, we also see a very faint galaxy 10\arcsec\ to the S
of the QSO. Thus, if there are companions, they are faint in the near
infrared.  Table 4 shows our photometry of the objects.

The QSO does show a faint straight `jet' or tail in both $J$ and $H$, extending
3\farcs5 to the N (see figure~\ref{fig2}).  The feature is almost
parallel to the diffraction spikes seen in the guide star, but is at
11\arcdeg\ to it, and emerges off-centered from the QSO. Thus, we
consider it is real, but it is difficult to measure accurately because
of its faintness, the varying sky near the guide star, and the presence
of diffraction spikes. Measures were made after surface fitting and
subtraction of the sky, and also after rotation and subtraction, both
in full resolution and 4 x 4 block-averaged images.

The jet is relatively featureless, but appears to have a small opening
angle. It suggests, once again, that there has been a recent tidal
event, even in this apparently empty field. Deep optical images may
show if there are companions with younger stellar populations present.
The color of the jet is comparable with that of the whole QSO, but its
value is poorly determined.

Aside from the jet, the QSO image is not detectably extended, and has
the same profile as the stars. However, the image resolution is lower
than average in these data (0\farcs24) due to the fainter guide star.

\section{1337$-$013}

  This is a radio-loud QSO at z = 1.62. The radio source is compact and
variable, with no published structure. Our data show the QSO to have 
at least 4 companions within 15\arcsec. The closest one marked in Figure 1
lies about 4\arcsec\ to the NE. However, there is also a resolved faint 
knot 2\arcsec\ S of the QSO (see
figure~\ref{fig3}) and possible faint flux extending to it. These
nearby features are seen more clearly in $J$ band, and they have remarkably
blue colors. Table 5 shows the measurements on the objects.

This QSO is another that appears to be in a group of galaxies, and 
interacting with one of its close neighbours. The blue color of the close
companions suggests that they are both dominated by active star-formation,
presumably as result of the tidal events with the QSO, or its nuclear
radiation: they are projected at 10 and 20 kpc from the QSO nucleus.

\section{1540+180}

This is a radio-loud quasar at z = 1.66 (4C18.43) listed with
approximate visual magnitude 18 by Hewett \& Burbidge (1993). The radio
structure is a bent FR II structure with bright knots along the lobes,
which suggests that there is interaction with dense and clumpy IGM. We
obtained images at $J$, $H$, and $K$, with good signal over
approximately 12\arcsec\ radius around the QSO. In these images
(Figure 5 shows the region near the QSO) the quasar is the brightest 
object, lying in a
uniformly populated region of eight galaxies which are up to 3 mag
fainter than the QSO itself.  It is possible that these are members of
a group associated with the quasar, from their high density and similar
colors. Table 6 shows photometry of several: the others lie too close
to the edge of the field to enable good measures.

Figure~\ref{fig4} shows the vicinity of the QSO.  The closest object
(to the E) seen clearly in $J$ and $H$ is an artifact produced by the
guide star: there are several of these in a characteristic pattern. The
$K$ band image shows a clear NE extension to the QSO and possible faint
luminosity further to the N. The $H$ band image does not show the
bright arm to the NE. However, there is a curved region of extended
luminosity extending about 4\arcsec\ to the N and ending in a brighter
region. In the $J$ band image this long curved region is only
marginally detected, but the bright region at the end is present. This
luminosity is in the general direction of the nearest galaxy but does
not connect with it, to the limit of our detection. We see no
luminosity to the S of the QSO.

The radio structure to the N matches the near infrared faint luminosity
quite closely, but not exactly. The radio lobe to the N has the same
length and lies mostly alongside, to the E of the near infrared flux,
and the edge-brightened lobe ends next to the brighter region of $H$
and $J$ emission. The radio structure suggests that the N lobe is the
more active at the present time. The $K$ band extension of the QSO does
not match the direction of the radio structure nearest the nucleus.

  This is the only field for which we have three colors, and are able
to make a color-color plot to compare with stellar population models.
Two-color photometry of the QSO and its companions is shown in
figure~\ref{fig5}, compared with age tracks for stellar populations at
different redshifts. Object 5 appears as a larger disk galaxy and in
this plot corresponds with a slightly reddened galaxy of intermediate
age at redshift $\sim$0.3. The other nearby galaxies lie in positions
on the plot that correspond to the QSO redshift. Another compact galaxy
near the edge of the field is not seen in all colors due to pointing
offsets, but may also be a companion. The QSO nucleus has the bluest
color, but the colors of the companions indicate they are all young,
with a spread of age or dust content. The objects are compact, with
diameters in our image from 2 to 5 kpc (at 5 kpc per arcsec: all are
smaller than one arcsec).  Thus, this appears to be a compact group of
galaxies dominated by young stellar populations. A similar situation is
seen around the z = 2.2 QSO 1345+580 (Hutchings 1998). The galaxies are
luminous and comparable with that of the QSO host, and lie within a
diameter of less than 100 kpc: thus this is a very dense group, which
is likely to evolve by merging or breaking up over cosmic time.

  The QSO has several lower redshift absorption line systems,
around z=0.7, and one at z=1.46. The companion galaxy colors do not
correspond to objects at z=0.7, but we cannot exclude their lying
at 1.46, which is close to the emission line redshift of 1.66.

  Crude PSF subtraction in $K$ band shows the NE tail clearly and indicates 
a resolved flux comparable to the companion galaxies (see
figure~\ref{fig6}).  The QSO is well resolved in this band and the
subtracted host galaxy flux is not signficiantly dependent on the PSF
correction for distance from the guide star. In $H$ band, however, the
resolution of the host galaxy is marginal and depends strongly on the
PSF correction. In the $K$ band resolved flux, photometric
K-corrections are likely to be no more than a few tenths (small if the
populations are young), so the luminosities are typical of luminous
galaxies.

  The brightest companion (\#1) has an exponential inner luminosity
  profile ($\sim$1 kpc) but a peculiar profile at larger radii. None of
the companions has any structure at near infrared wavelengths - they
are smooth and symmetrical and well-resolved, even though their colors
suggest young stellar populations. They have bright compact nuclei.

\section{0849+120}

   This QSO is faint at v = 20.5, and is radio-quiet. It lies
in a field with several other objects with a wide range of brightness and
size. Only two of these are likely to be associated with the QSO 
(see figure~\ref{fig7}): the others are too bright and/or too large.

  The closest companion (\#1) lies about 5\arcsec\ to the N, and 
slightly E, and in the $K$ band
image there may be a faint connecting luminous bridge between it and the QSO.
However, this bridge is not seen in $H$ band and there are sky background fluctuations
which are comparable. The QSO itself is not measurably extended.

\section{1236$-$003}

This QSO is at z = 2.18, is optically moderate at v = 19.1, and is radio-quiet.
It has three companions all at about 15\arcsec\ separation and well
separated in direction. Table 8 shows their magnitudes, which are too
bright to be likely associated galaxies. Their sizes and colors are also 
consistent with their being at considerably lower redshift - perhaps 0.2.

The closest companion to 1236$-$003 is seen most clearly in the
$H$ band image, at some 4\farcs3 to the NW. 
It is also (just) detected in the $K$ band image. There is no detectable flux
connection to the QSO, or tidal elongation. In both H and K band, the
QSO images appear to have extended flux to the NE, curving to the E.
This is faint enough to be questionable, but
suggestive as it appears identically in both H and K. While colors are
subject to errors at these faint levels, the companion is blue
and the extended flux is redder than the QSO, as given in Table 8.

The structure and companion seen here are fainter than in the radio-loud 
objects, and
the field is otherwise lacking in companions  as bright as seen in the 
radio-loud objects. However, the data are suggestive of some tidal event 
in 1236$-$003 too.

\section{0552+398}

   This is a bright radio-loud QSO at z = 2.36. It lies in a very crowded
field, in which our $K$ band image clearly shows 20 objects besides the QSO.
Of these, 10 are bright or large enough that they cannot be companions.
Nine of the other ten objects form a band on either side of the QSO, some
5\arcsec\ wide and at least the length of the image -- 36\arcsec. These
objects are compact and in the range 4 to 6 magnitudes fainter than
the QSO. The table shows their magnitudes and also $H$ band absolute
magnitudes assuming they have standard colors ($H-K$ = 1) for a young
population at this redshift. We have omitted the galaxies that are too bright
to be companions for brevity in the table. Figure 1 shows the positions of
most of them - a few lie outside the boundary shown because of the dither.
The galaxies appear more luminous than those in other fields - possibly 
because there are significant negative K-corrections at this higher redshift, 
as would be the case for young populations: as in all cases, we
have adopted no k-correction. Also, H$\alpha$ is shifted into the $K$ band
at this redshift, which might increase the flux if these galaxies have
Balmer emission.

   The nearest companion lies about 2\farcs9  to the N and there appears
to be curved connecting luminosity to the QSO, suggesting tidal interaction
(see figure~\ref{fig9}).
Thus, we have evidence that the QSO is a member of a dense group
of compact galaxies of comparable luminosity to the host and each other.
The FWHM of the companions is 0\farcs35, and all are round with no
structure and a bright nucleus.

  The radio source is compact, luminous, variable, GHz-peaked, and has 
structure seen only with VLBI: a weak halo extended more E-W than N-S. 
This suggests the source is young.

\section{1307+297}

   This is a radio-quiet QSO at redshift 3.09. Our exposures are shorter than
for the other objects by a factor 3 to 4 so that our detection limits are
significantly different. In addition, it has a very bright guide star
which gives a high level of scattered light over much of the 36" field.
Nevertheless, we easily see one galaxy 11" away in $H$-band at $H$=20.3 
and can detect it at $K$=20.5, but no other objects. The blue color and
brightness indicate it is likely to be a foreground object. The QSO is
measured at $H$=19.1 and $K$=18.1, so its catalogue value of v=18.6 seems too
bright, or it is currently in a faint state.

   Given the poorer detection threshold and the bright guide star, our
observation of this
radio-quiet object provides weak evidence that it is not in a dense galaxy environment. The QSO itself does not appear to be resolved in these exposures.

\section{0104+0215}

This QSO is radio-quiet and has redshift 4.16. Because of its high redshift
we exposed much longer, and mainly in K-band (rest frame U/B)
on this field. The field 
appears empty and has one other object, of brightness similar to the QSO,
that we detect with similar limits to our other fields. However, when we
compare combined integrations of about 11000 secs each, from two separate
nights, some very faint objects and features are seen in both. Table 10
shows the measures of these and their positions are shown in Figure~\ref{map}.
Figure~\ref{fig10} shows the environment of the QSO.

  The objects are fainter than any others measured, and some are very
compact, while others are larger and diffuse. 
The group 3, 4, 5 are all diffuse and have
diameters of 2" to our limit of detection. The compact objects are
all 0\farcs4 to 0\farcs6 across. At the QSO redshift they have absolute $K$
magnitudes of $\sim$-25 without k-correction.

  The QSO itself appears to be extended in a near E-W direction, 
which is almost
perpendicular to the radius vector to the guide star, and so not likely to
be due to PSF effects, which are radial. The extended light
is stronger on the W side. In addition, there is a suggestion of a faint
smooth tail to the N and curving to the E, as seen in the figure. This
is so faint that it may be an artifact, and is seen more clearly on one
night's image than the other. Seeing and transparency were similar on the two
nights. The shorter $H$-band exposure does not reveal any of these objects
or features.

   We will discuss the PSF removal of this (and other) objects more
rigorously in a separate publication. Nevertheless, we have suggestive
evidence from our $K$-band data that this high redshift object has an
asymmetrical host galaxy, some nearby companions, and maybe a tidal tail
as well. The QSO magnitude is 18.3 in $H$ and 18.0 in $K$, giving it a 
very blue color.

\section{Discussion}

    Our relatively small QSO sample is fairly well-matched
between radio-loud and radio-quiet at redshifts near 1, 1.6, and
2.2. The small pixel size has required long integration times, so that
our images are limited by sky noise. The AO camera performance has
produced images with FWHM well under 0\farcs2 in most bands and
objects: we were careful to observe only those objects which have guide
stars close and bright enough to yield optimal correction. We have
reported only data obtained in good natural seeing and clear
conditions. Thus, the data on all objects are very comparable.

   The extended features seen in most objects are faint, so that we may
be missing some fainter structure that is lost in the sky noise.
This is increasingly true as we move through the $J$, $H$, $K$ range,
so that the faintest structures are seen only in $J$ band in all 3 cases 
where we observed in $J$. It appears
that there is no very small-scale structure present in any of the objects, 
so we binned the
data 4 $\times$ 4 (to effective pixels 0\farcs14) to decrease the sky
noise and make some measurements easier.

   It is notable that we find faint structures in all
observed objects to redshift 2.4, and probably also in the z=4.1 object.
These structures are smooth in overall shape and brightness,
and characteristic of tidal tails of stars from a galaxy, rather than
blue and knotty structure characteristic of active star formation.
However, our observations are at relatively long restframe wavelengths
so we are not very sensitive to star-formation activity. It will be
important to obtain shorter wavelength (i.e. CCD) data of high
resolution to check our conclusion that tidal activity is more
important than star formation. In any event, the data suggests that
tidal events are an even more marked cause (or effect) of QSO activity
at higher redshifts, than the well-known observations of interactions
at lower redshifts.

   In our sample to date, there is also a notable difference in the galaxy
environment between
radio-loud and radio-quiet QSOs, with the latter generally being located
in poorer environments. The exceptions are the radio-loud but apparently 
isolated QSO 0804+499, and possibly the radio-quiet QSO 0104+022, 
although the companions to the
latter are detected only because of our much deeper exposure in this field. 
We caution against a general conclusion with our small and incomplete sample,
but note that this environmental difference is an extension of what is
seen at lower redshifts, although we are only able to detect the
brightest of possible companion galaxies. We note also that at
visible wavelengths, Hutchings (1995) finds excess galaxies around
both radio-loud and radio quiet QSOs. Thus, the difference may possibly be
in the population age of companions. The space density of the NIR-observed
companions to the radio-loud QSOs is very high - on average the
separation between galaxies in the four crowded fields observed is 35
kpc.  This suggests that merging will be an important part of the
activity in these groups with time. Until we can study the environments
to greater depth, we cannot tell if these are simply small groups or
may be the cores of what will evolve to major cluster at lower
redshift, by accretion from the field. Here too, deep CCD observations
with high resolution are called for, to take advantage of the lower sky
brightness.

  One possible related issue is the connection between objects in the
field and foreground absorbers. The Hewitt Burbidge (1993) catalogue has
references to absorbers in only one object: 1540+180. Thus, this is the only
known case where we may have likely foreground confusion. (However, it
may be that other QSOs have not been observed with sufficient resolution
to detect foreground absorbers.) We have noted
in the section on 1540+180 that we may be seeing absorbing objects from
the z=1.46 system, but not the others at z$\sim$0.7. At this point, our main
inferences about the QSO environments are not affected by known foreground
absorbers.

   Five of our sources are radio-loud. Two have no published radio maps
(but are almost certainly compact) and two others are unresolved in the
radio.  The only source with comparably-sized resolved radio structure is
1540+180, which has some (but not all) corresponding  visible/near
infrared structure. In this connection we also note the radio-loud QSO
1345+580 (Hutchings 1998) has L$\alpha$ emission at the places of
bright radio knots in a very bent radio structure.

   In the only field where we have two colors, and hence can estimate 
redshifts, we find that there are
several companions that appear to lie at the QSO redshift. In other
fields, we eliminate foreground objects by their larger size and scale
length, and in many cases their brightness. The remaining, possible 
companion galaxies are all much more compact -
less than one arcsec (5 kpc) in diameter. They are mostly round and
featureless except for unresolved nuclei in some. For galaxies where
the signal-to-noise is sufficient to measure luminosity profiles they
are neither pure disk nor power law.  The small size  is consistent
with reports for other high redshift galaxies (e.g. Steidel et al.
1996).  The smooth round morphologies suggest that the older stellar
populations that we are imaging have regular shapes, in contrast with
more irregular shapes seen in visible (rest UV) wavelengths, where we
are seeing the clumpy nature of star-forming regions and, possibly,
results of dust obscuration.  The small size and relatively high
luminosity of these objects suggests that they represent a stage of the
early formation of galaxies. The high space density in some fields also
suggests that these galaxies may merge with time, to form present day
large galaxies. A similar scenario has been proposed by Rottgering(1997) for
the dense environments around high redshift radio galaxies.
   
\newpage

\begin{table} 
\begin{center}
\caption{Table of near infrared imaging results - ordered by redshift}
\vskip 0.3cm
\begin{tabular}{l|l|l|l|l}
\hline
QSO &m$_v$ &z &Radio &Comment\\
\hline
\smallskip
0915$-$213 &17.5 &0.85 &RL &Resolved host, knot, tail, compact cluster\\
1055+019 &20.5 &1.06 &RQ &Resolved, faint jet or extended companion *\\
0804+499 &17.5 &1.43 &RL &$\sim$Unresolved, tail \\
1337$-$013 &18.7 &1.61 &RL &Compact group, poss connection to nearest knot\\
1540+180 &18 &1.66 &RL & Resolved, tail, compact cluster\\
0849+120 &20.5 &1.76 &RQ &2 possible companions, maybe tail to one\\
1236$-$003 &19.1 &2.18 &RQ &$\sim$Unresolved, poss tail \& nearby companion\\
0552+398 &18 &2.37 &RL &Interacting? dense cluster\\
1307+297 &18.6 &3.09 &RQ &Short exposures, unresolved, one companion\\
0104+022 &19.7 &4.16  &RQ &Resolved + tail? faint companions\\
\hline
\smallskip
* See Hutchings et al 1998
\end{tabular}
\end{center}
\end{table}

\begin{deluxetable}{llllrccccc} 
\tablenum{2}
\tablecaption{Journal of observations - ordered by RA}
\tablehead{\colhead{Name} &\colhead{Month} 
&\multicolumn{3}{c}{Exposure in secs} 
&\colhead{Guide * mag} &\colhead{Offset} &\multicolumn{3}{c}{FWHM (\arcsec)}\\
&1998 &\colhead{$J$} &\colhead{$H$} &\colhead{$K$} &\colhead{$R$} &\colhead{arcsec}
&\colhead{$J$} &\colhead{$H$} &\colhead{$K$}}
\startdata
0104+022 &Jan &-- &4800 &24000 &13.8 &14 &-- &0.32 &0.17\nl
0552+398 &Jan &-- &- &9300 &11.7 &26 &-- &-- &0.24\nl
0804+499 &Mar &6000 &7200 &- &14.6 &11 &0.23 &0.24 &--\nl
0849+120 &Jan &-- &9600 &7200 &13.4 &16  &-- &0.16 &0.19\nl
0915$-$213 &Mar &-- &7200 &7200 &13.2 &12 &-- &0.16 &0.16\nl
1236$-$003 &Jan &-- &9600 &7200 &12.8 &19  &-- &0.25 &0.15\nl
1307+297 &Jan &-- &1500 &1800 &8.1 &21 &-- &0.15 &0.15\nl
1337$-$013 &Mar &7200 &7800 &-- &13.4 &20 &0.19 &0.15 &--\nl
1540+180 &Mar &6000 &6000 &6000 &12.6 &17 &0.23 &0.14 &0.13\nl
\enddata
\end{deluxetable}

\begin{deluxetable}{lllll} 
\tablenum{3}
\tablecaption{Photometry in 0915$-$213 field}
\tablehead{\colhead{Object} &\colhead{$K$} &\colhead{$H$}  
&\colhead{$H-K$} &\colhead{M$_H$}}
\startdata
Q &15.9 &16.7  &0.8  &$-$25.8 \nl
Qhost &20.3 &20.6  &0.3  &$-$21.9 \nl
1 &17.9 &18.0 &0.1 &star\nl
2 &18.7 &19.1 &0.4 &$-$23.4\nl
3 &21.0 &21.7 &0.7 &$-$20.8\nl
4 &- &19.5  &- &$-$23.0\nl
5 &19.1 &19.8 &0.7  &$-$22.7\nl
wisp &20.5 &21.4  &0.9 &$-$21.1 \nl
\enddata
\end{deluxetable}

\begin{deluxetable}{lllll} 
\tablenum{4}
\tablecaption{Photometry in 0804+499 field}
\tablehead{\colhead{Object} &\colhead{$J$} &\colhead{$H$}  
&\colhead{$J-H$} &\colhead{M$_H$}}
\startdata
Q &16.9 &15.9  &1.0  &$-$27.0 \nl
Qjet &22.5 &21.9  &0.6:  &$-$21.4 \nl
star &17.1 &16.6 &0.5 &- - -\nl
1 &19.2 &18.7 &0.5 &$-$24.7\nl
\enddata
\end{deluxetable}

\begin{deluxetable}{lllll} 
\tablenum{5}
\tablecaption{Photometry in 1337$-$013 field}
\tablehead{\colhead{Object} &\colhead{$H$} &\colhead{$J$}  
&\colhead{$J-H$} &\colhead{M$_J$}}
\startdata
Q &17.3 &18.1  &0.8 &$-$26.2 \nl
nearby knot &22.3 &22.3  &0.0 &$-$21.9\nl
1 &22.5 &22.4  &$-$0.1 &$-$22.0\nl
2 &20.4 &21.2  &0.8  &$-$23.1\nl
3 &&22.7\nl
4 &&22.8\nl
\enddata
\end{deluxetable}
   
\begin{deluxetable}{lllllll} 
\tablenum{6}
\tablecaption{Photometry in 1540+180 field}
\tablehead{\colhead{Object} &\colhead{$K$} &\colhead{$H$} &\colhead{$J$} 
&\colhead{$J-H$} &\colhead{$H-K$} &\colhead{M$_J$}}
\startdata
Q &17.6 &18.3 &19.0 &0.7 &0.7 &$-$25.4 \nl
Qhost &19.2 \nl
1 &18.7 &19.6 &20.3 &0.7 &0.9 &$-$24.1\nl
2 &19.7 &20.7 &21.6 &0.9 &1.0 &$-$22.8\nl
3 &19.4 &20.6 &21.7 &1.2 &1.2 &$-$22.9\nl
4 &20.5 &21.4 &22.0 &0.6 &0.9  &$-$22.6\nl
5 &18.5 &19.0 &20.0 &1.0 &0.5 \nl
\enddata
\end{deluxetable}

\begin{deluxetable}{llllll} 
\tablenum{7}
\tablecaption{Photometry in 0849+120 field}
\tablehead{\colhead{Object} &\colhead{$H$} &\colhead{$K$}  
&\colhead{$H-K$} &\colhead{M$_H$} }
\startdata
Q &19.5 &19.1  &0.4 &$-$25.0 \nl
1 &22.2 &21.3  &0.9 &$-$22.3 &Compact\nl
2 &18.8 &18.5  &0.3 &&Large\nl
3 &22.2 &-  &- &&Large faint\nl
4 &22.1 &-     &- &$-$22.4 &Compact\nl
5 &19.7 &19.6 &0.1 &&Bright\nl
6 &17.5 &17.3 &0.3 &&Star\nl
\enddata
\end{deluxetable}

\begin{deluxetable}{lllll} 
\tablenum{8}
\tablecaption{Photometry in 1236$-$003 field}
\tablehead{\colhead{Object} &\colhead{$H$} &\colhead{$K$}  
&\colhead{$H-K$} &\colhead{M$_H$}}
\startdata
Q &17.4 &16.7  &0.7 &$-$27.9 \nl
1 &19.4 &19.2  &0.2 \nl
2 &18.3 &17.9  &0.4 \nl
3 &20.3 &19.8  &0.5 \nl
4 &22.0 &21.9     &0.1 &$-$23.3\nl
QSO ext &21.0    &20.0 &1.0 &$-$22.3 \nl
\enddata
\end{deluxetable}

\begin{deluxetable}{lll} 
\tablewidth{4cm}
\tablenum{9}
\tablecaption{Photometry in 0552+398 field}
\tablehead{\colhead{Object} &\colhead{$K$} &\colhead{M$_H$}}
\startdata
Q &15.2  &$-$29.3\nl
1 &19.1 &$-$25.4\nl
2 &20.7  &$-$23.8\nl
3 &20.9  &$-$23.6\nl
4 &20.5  &$-$24.0\nl
5 &19.3 \nl
9 &19.7 &$-$24.8\nl
10 &19.7 &$-$24.8\nl
13 &20.4 &$-$24.1\nl
14 &20.7 &$-$23.8\nl
17 &20.6 &$-$23.9\nl
18 &20.4 &$-$24.1\nl
\enddata
\end{deluxetable}

\begin{deluxetable}{lll} 
\tablewidth{3cm}
\tablenum{10}
\tablecaption{Photometry in 0104+022 field}
\tablehead{\colhead{Object} &\colhead{$K$} }
\startdata
Q &18.0\nl
1 &21.8\nl
2 &22.2\nl
3 (large) &20.6\nl
4 (large) &20.9\nl
5 (large) &21.4\nl
6 (star?) &18.7\nl
7 &21.7\nl
8 &21.8\nl
9 &20.9\nl
\enddata
\end{deluxetable}

\centerline{References}

Aretxaga I., Boyle B.J., Terlevich R.J., 1995, MNRAS, 275, 27

Aretxaga I., Le Mignant D., Melnick J., Terlevich R.J., Boyle B.J., 1998,
MNRAS, 298, L13

Bahcall J.N., Kirhakos S., Saxe D.H., Schneider D.P., 1997, \apj, 479, 642

Campos A. et al 1998 1998 ApJL, in press (astro-ph/9809146)

Casali, M. \& Hawarden, T. 1992, JCMT-UKIRT Newsletter, 4, 33

Heckman T.M., Lehnert M.D., van Breugel W., Miley G.K., 1991, \apj, 370, 78

Hewitt, A. \& Burbidge, G. 1993, Ap. J. Suppl., 87, 451

Hutchings J.B. \& Neff S.G., 1992, AJ, 104, 1

Hutchings J.B., Holtzman J., Sparks W.B., Morris S.C., Hanisch R.J.,
Mo J., 1994, ApJ, 429, L1

Hutchings J.B. 1995, AJ, 109, 928

Hutchings J.B. \& Neff S.G. 1997a, AJ, 113, 550

Hutchings J.B., \& Neff S.G., 1997b, AJ, 113, 1514

Hutchings J.B. 1998, AJ, 116, 20

Hutchings J.B., Crampton D., Morris S.L., Steinbring E., 1998, PASP, 110, 374

Lehnert M.D., Hackman T.M., Chambers K.C., Miley G.K., 1992, \apj, 393, 68

McLeod K. K. \& Rieke G. H., 1995, \apj, 454, L77

Pentericci L., et al 1998, AAp, in press (astro-ph/980956)

Rigaut, F. et al. 1998, PASP, 110, 152

Rottgering, H. 1997, IAU Symp., 186, 184

Steidel, C.C., Giavalisco, M., Pettini, M., Dickinson, M. \& Adelberger, K.L.
1996, \apj, 462, L17

\clearpage

\centerline{\bf Captions to figures}

\figcaption[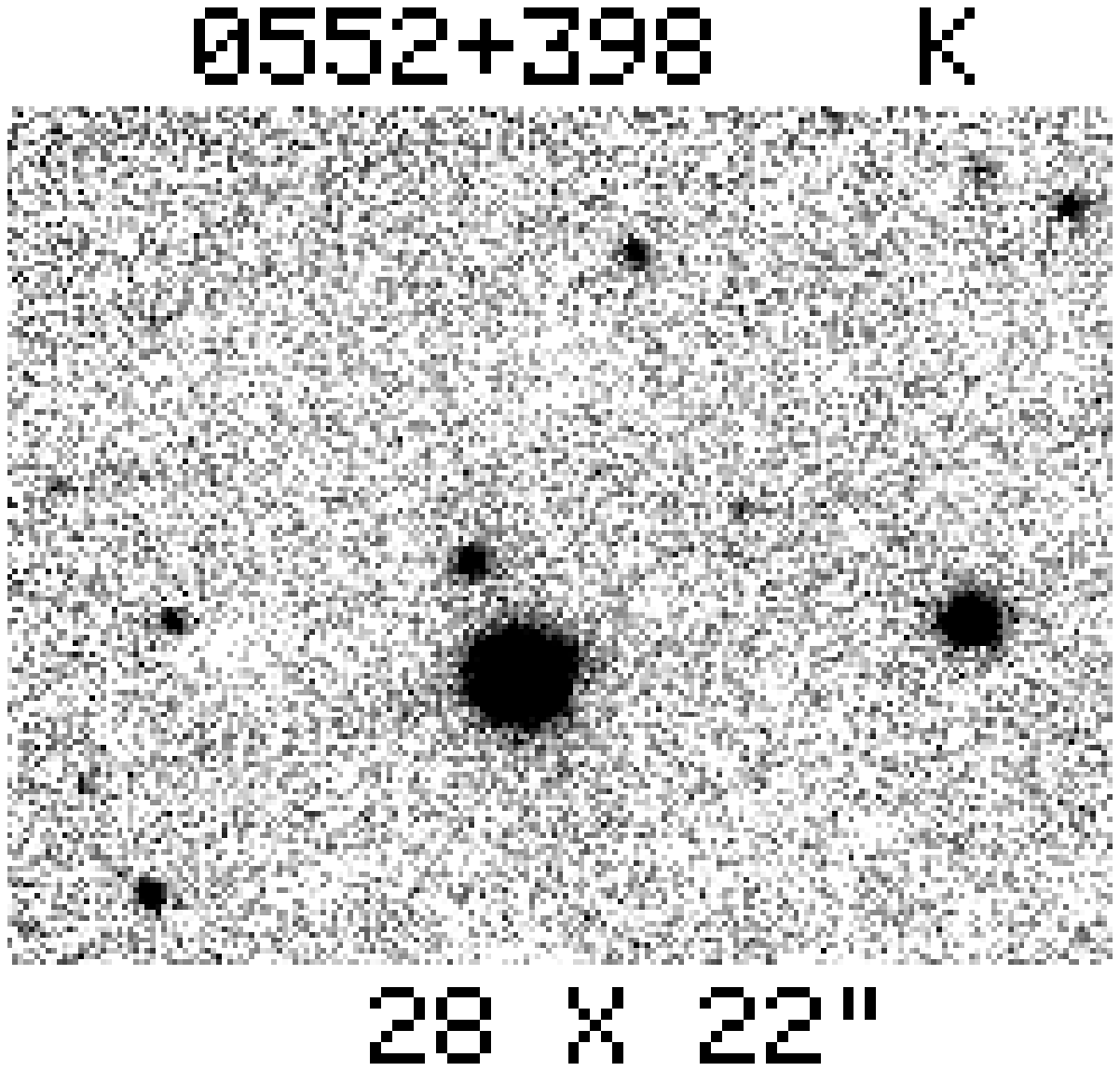]{Location of detected objects in all fields.
The field side is 36\arcsec. N is up and E to the left. Some objects not shown
were seen in partially exposed larger field due to dither patterns.
The guide star is indicated as GS when it is in the field. The QSO is 
indicated by Q, assumed foreground galaxies by F, and stars by *.
Photometry of most numbered galaxies is in the Tables. 
\label{map}}

\figcaption[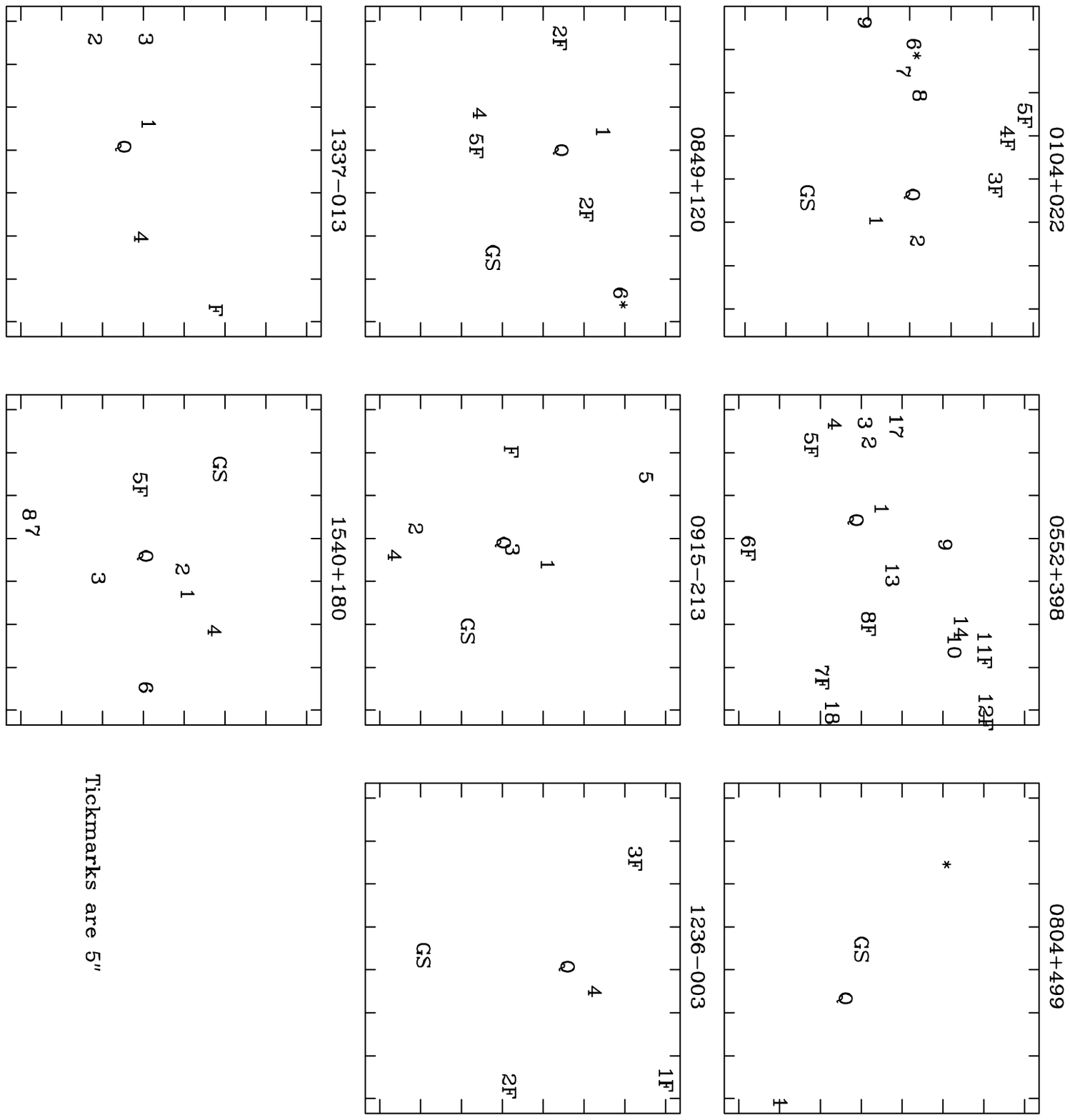]{$K$ and $H$ band images of 0915$-$213
(lower bright object). The images have been binned 4 $\times$ 4 (to
0\farcs14 pixels) to enhance faint features. N is up and E to the left.
Note the extension of the QSO to the NE and the nearby compact object
to the NW. There is probably a faint arc of flux to the NW, seen in both 
bands, but more clearly in K. \label{fig1}}

\figcaption[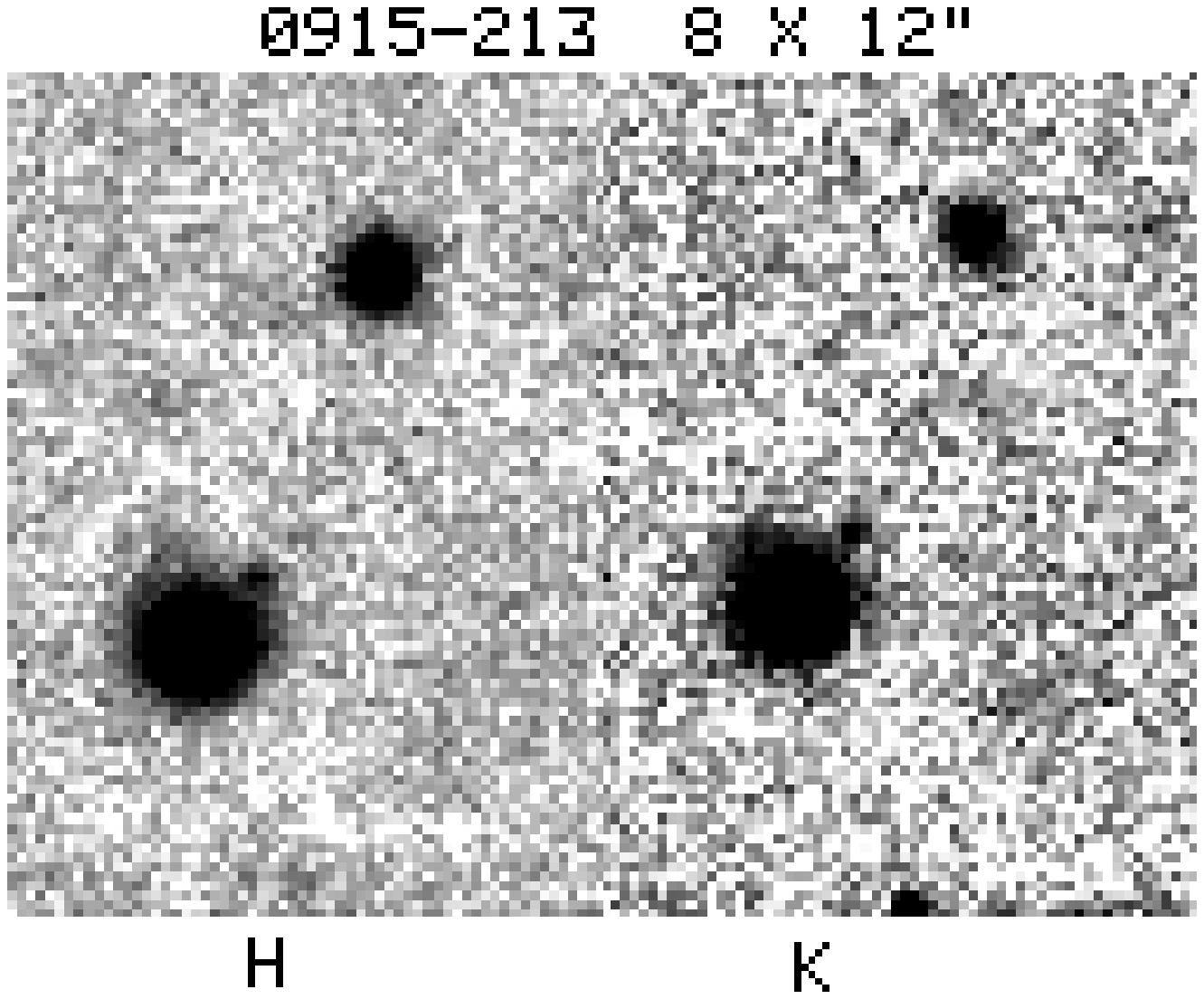]{J band image of 0804+499 with full
resolution and with 4 $\times$ 4 pixel binning to enhance visibilty of
faint jet. \label{fig2}}

\figcaption[hutchings.figure3.ps]{$J$ and $H$ band images of
1337$-$013, showing companion to NE and closer companion to the S,
possibly with connecting flux. The greater visibility in $J$ band may be 
due to redshifted [O~III] emission. 
\label{fig3}}

\figcaption[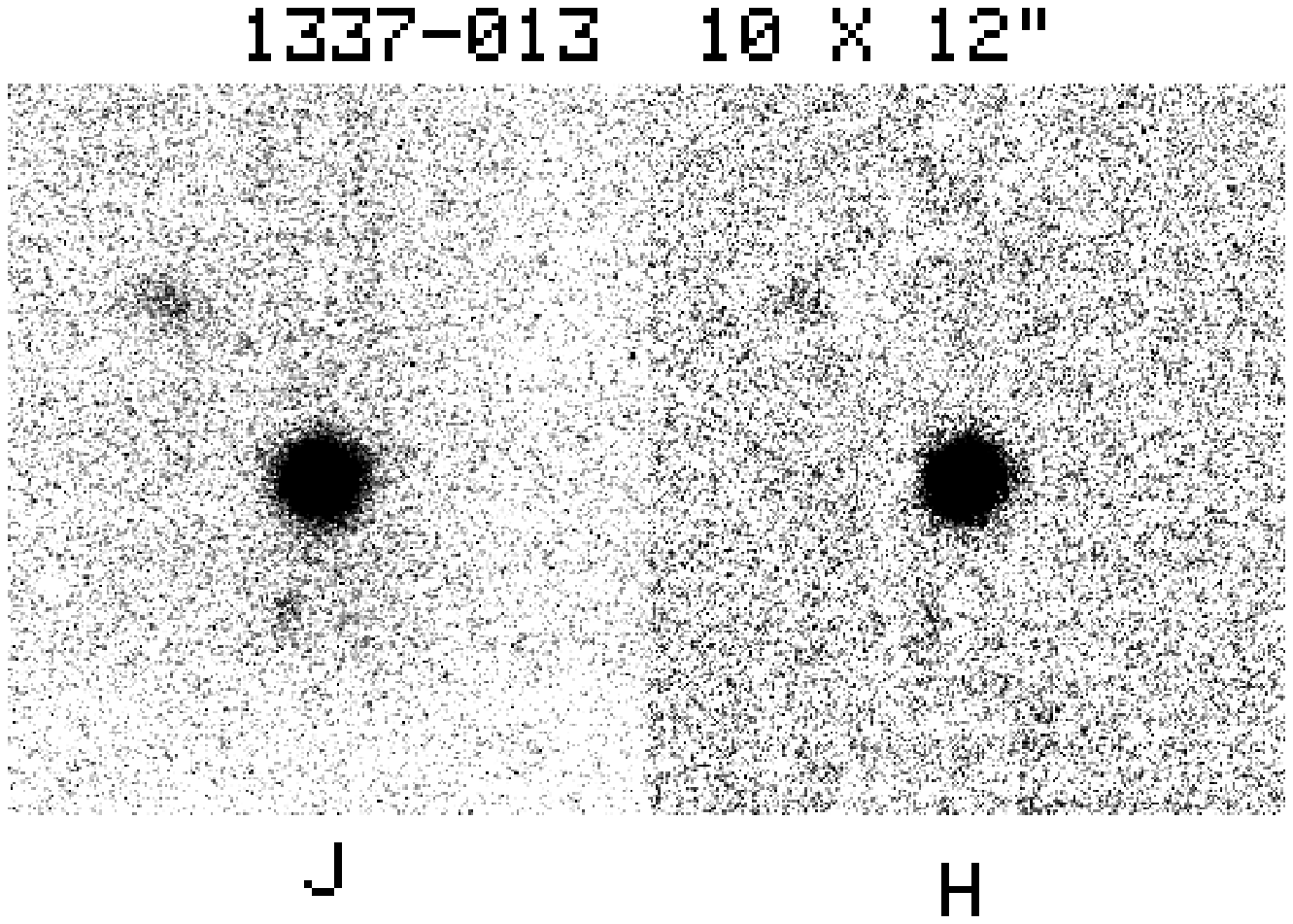]{4 x 4 binned images in 3 bands around 
1540+180. The
QSO is extended to the NE, particularly in $K$ band. In the H-band this
extends to a long faint wisp terminating at a round blob, which is
visible in all three bands. The radio structure to the N
lies close to this near infrared wisp. \label{fig4}}

\figcaption[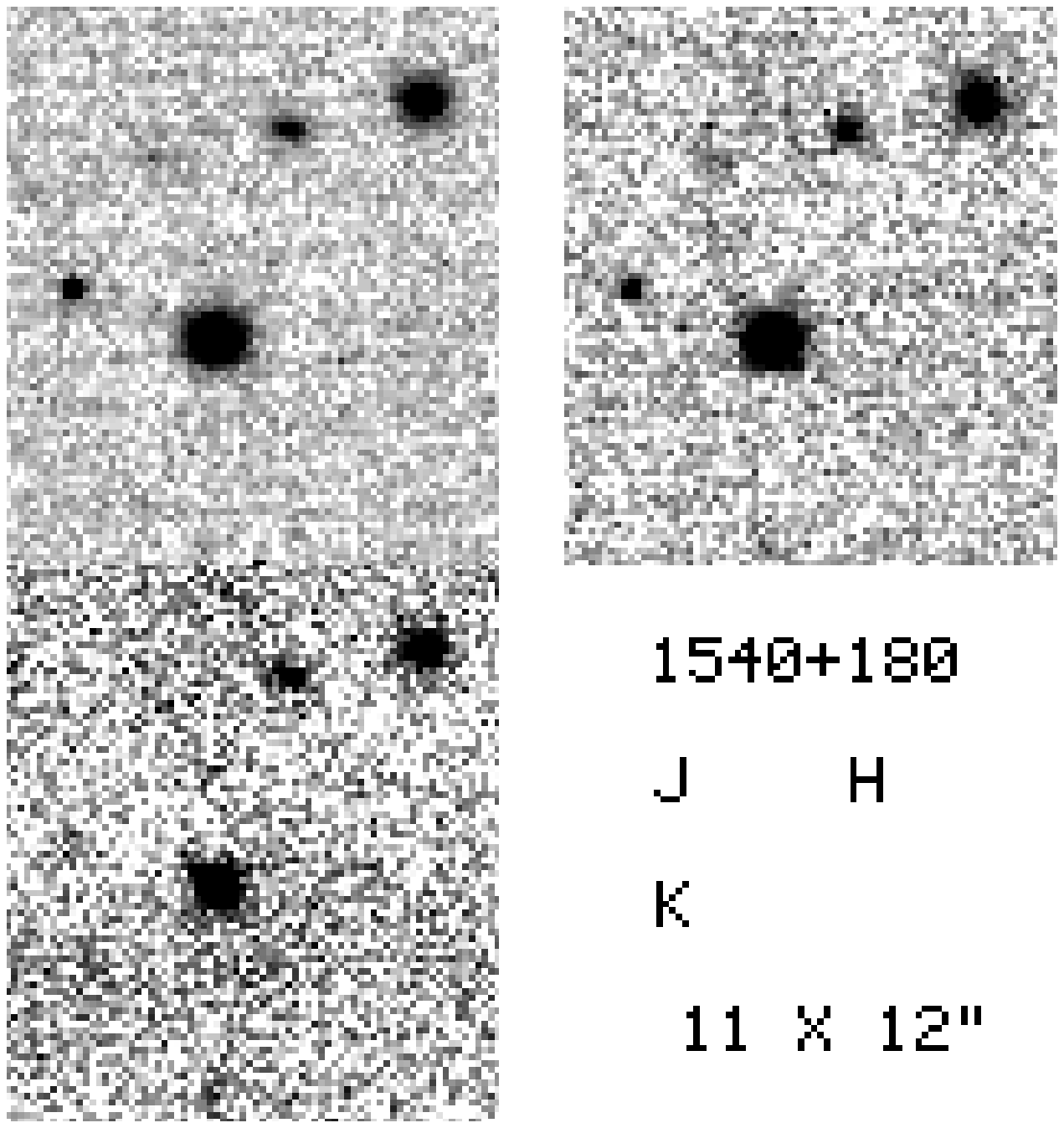]{near infrared 2-color plot showing
population evolution after a 1 Gyr starburst, at different redshifts.
Symbols along the tracks mark the ages indicated, and the point below 1
is 0.1 Gyr. The dotted line shows the effect of extinction on the
z=1 track. 
The QSO and companions 1, 2, 3, and 4 lie close to the positions
expected for young propulations at the QSO redshift of 1.7, possibly
with some dust. Object 5 is probably a reddened foreground galaxy at
z$<$0.3. \label{fig5}}

\figcaption[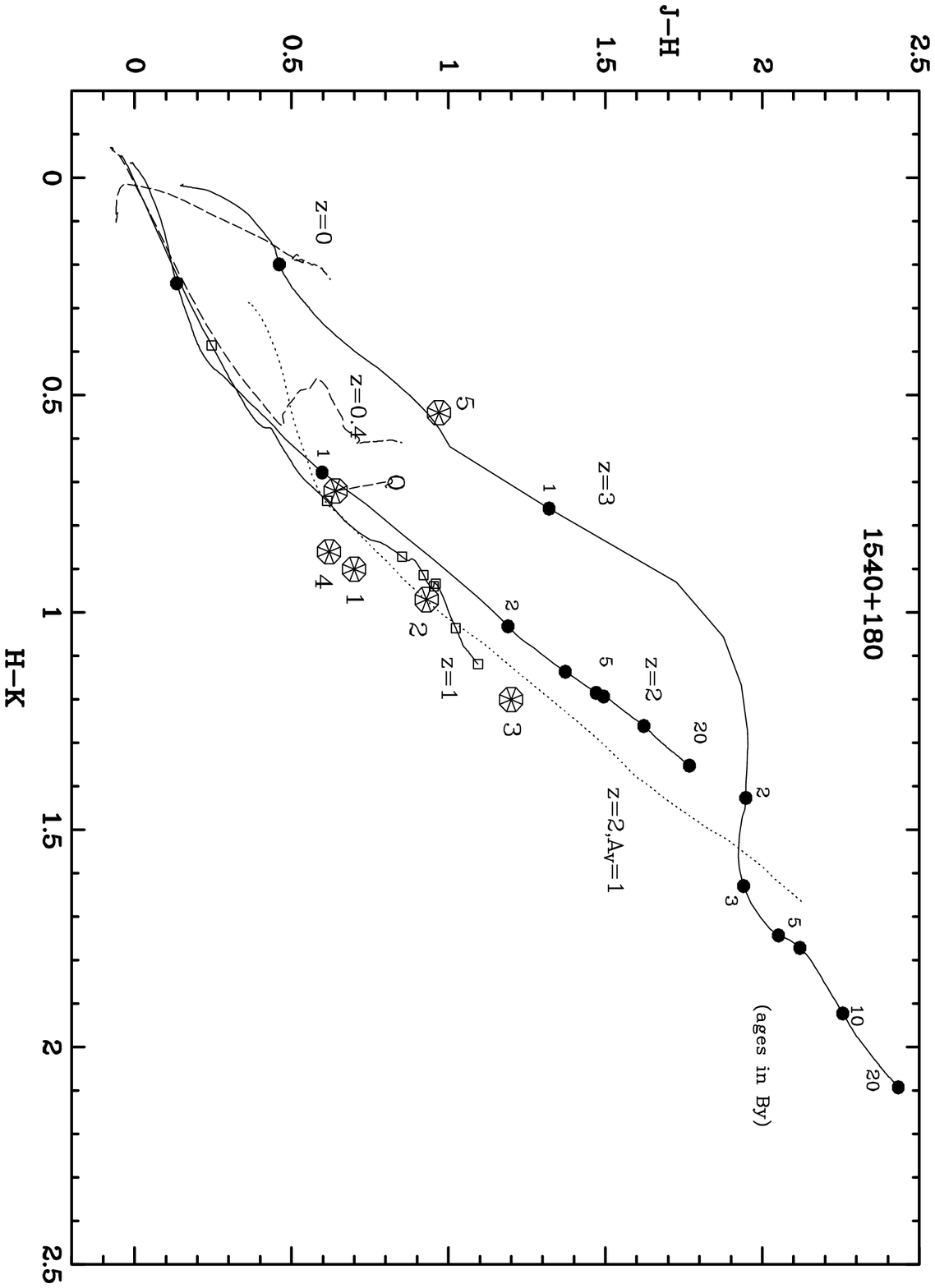]{Azimuthally averaged profiles of QSO
1540+180 and the PSF, approximately corrected for distance from
the guide star. \label{fig6}}

\figcaption[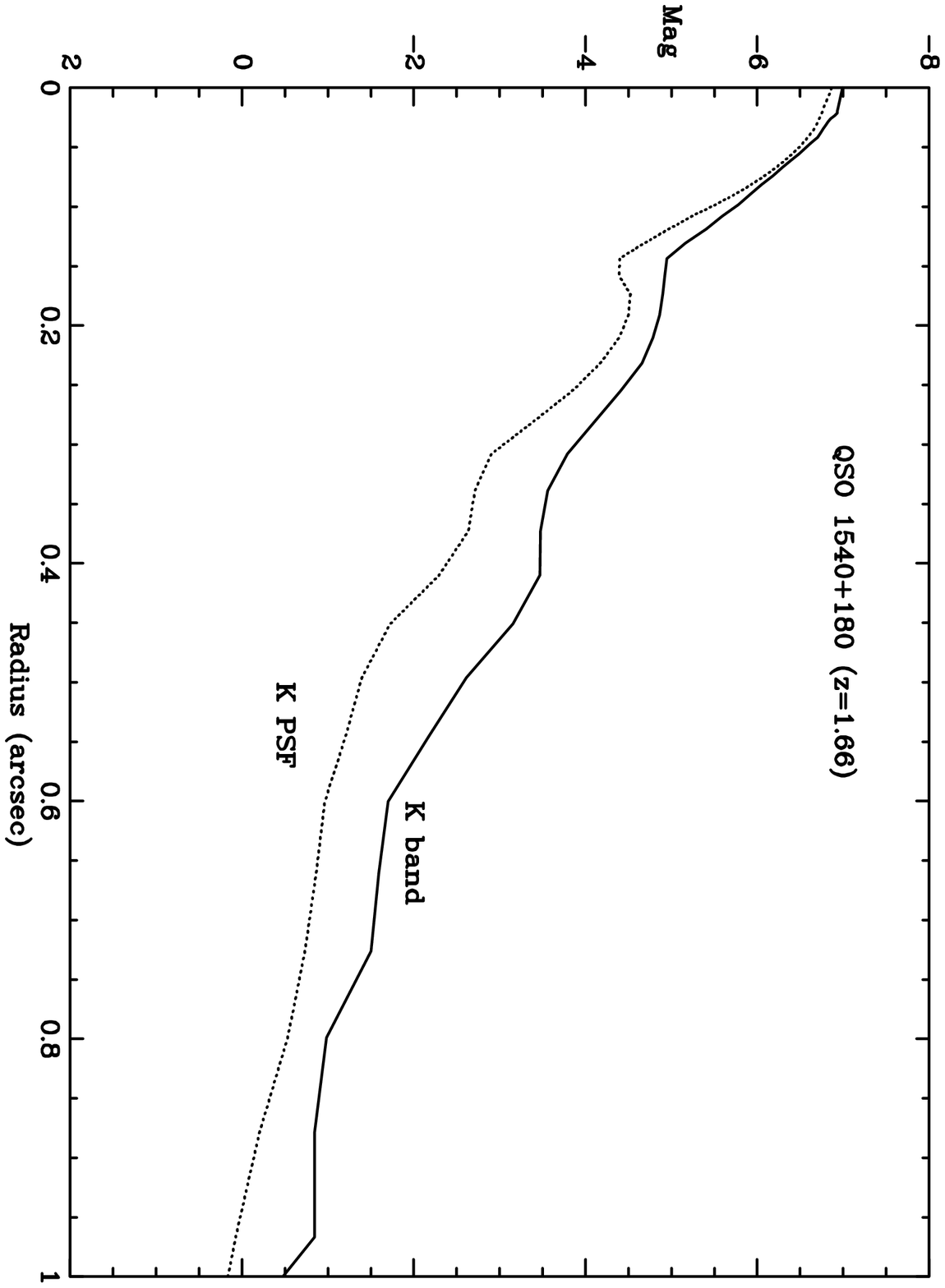]{$H$ and $K$ band images of
the field around QSO 0849+120, binned 4 $\times$ 4. In the $K$ band 
there is a luminous connection to the nearest companion, to the NE. \label{fig7}}

\figcaption[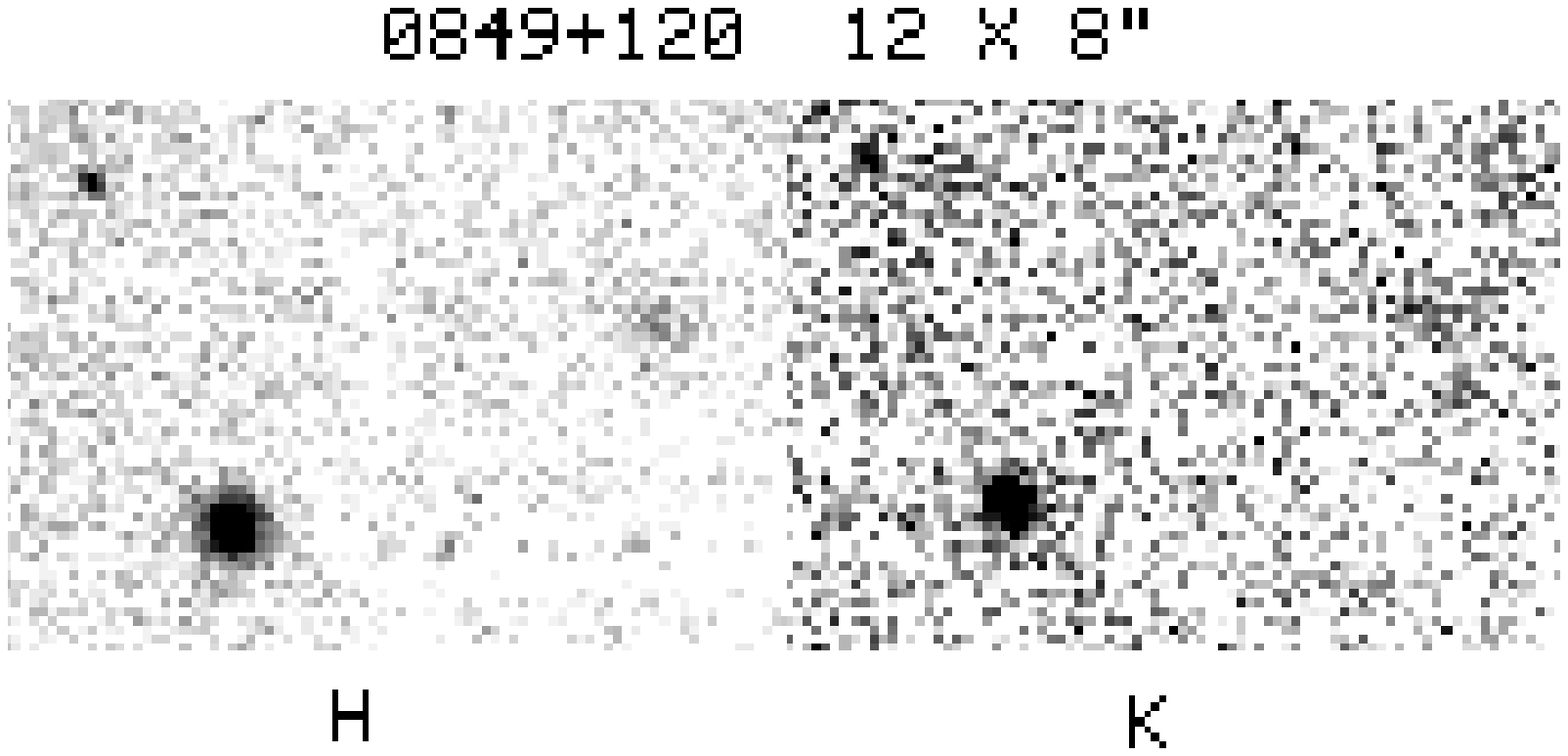]{$H$ and $K$ band images, binned 4
$\times$ 4, of QSO 1236$-$003. There is a compact companion to the NW
(seen more clearly in H), and connected extended flux to the NE (seen 
similarly in both H ans K). \label{fig8}}

\figcaption[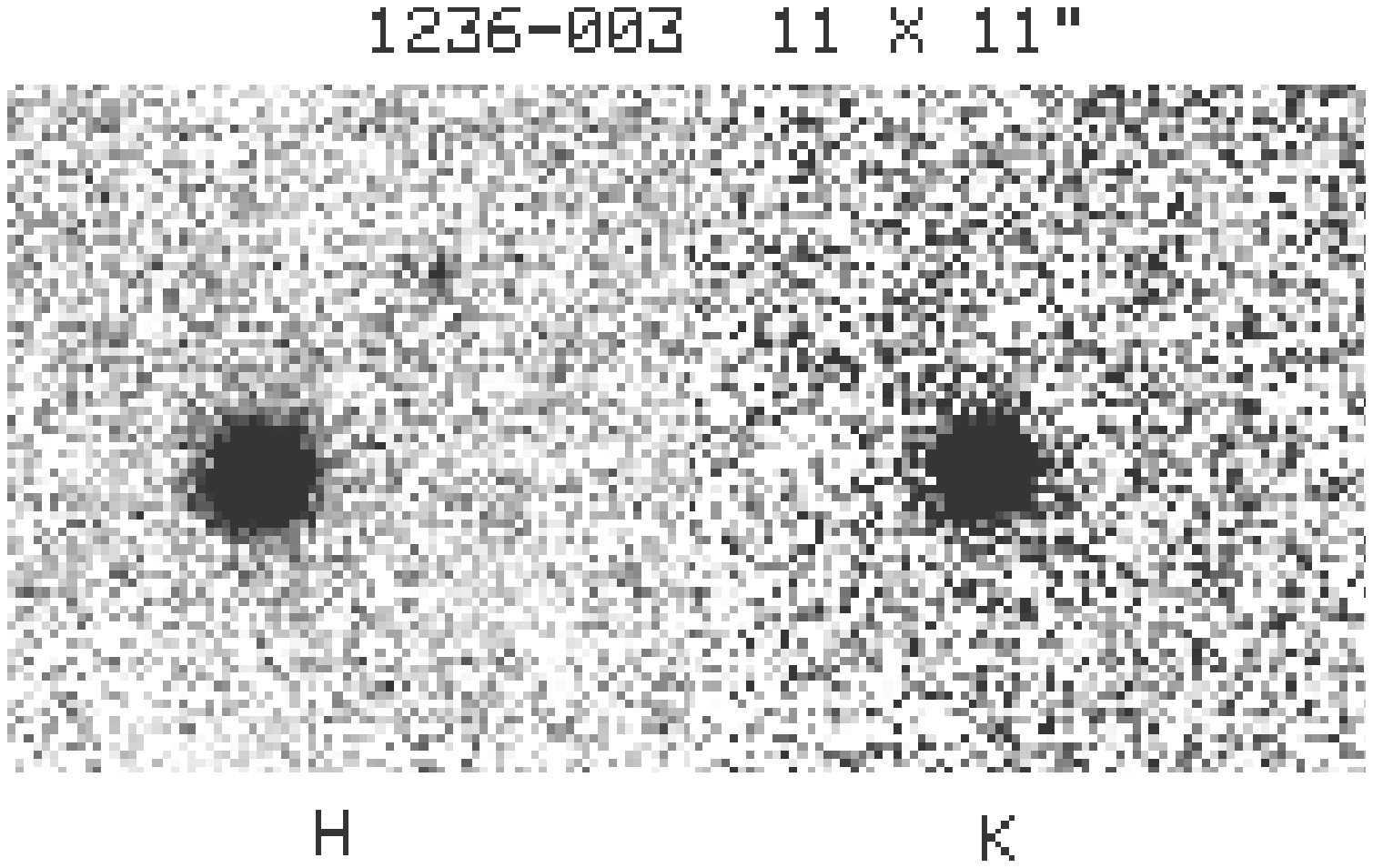]{$K$ band image, binned 4 $\times$ 4 of
4\arcsec\ $\times$ 3\arcsec\ field around QSO 0552+398. There is probably
connecting flux to the compact companion to the N. Most of the other objects 
are thought to be compact companions to the QSO. \label{fig9}}

\figcaption[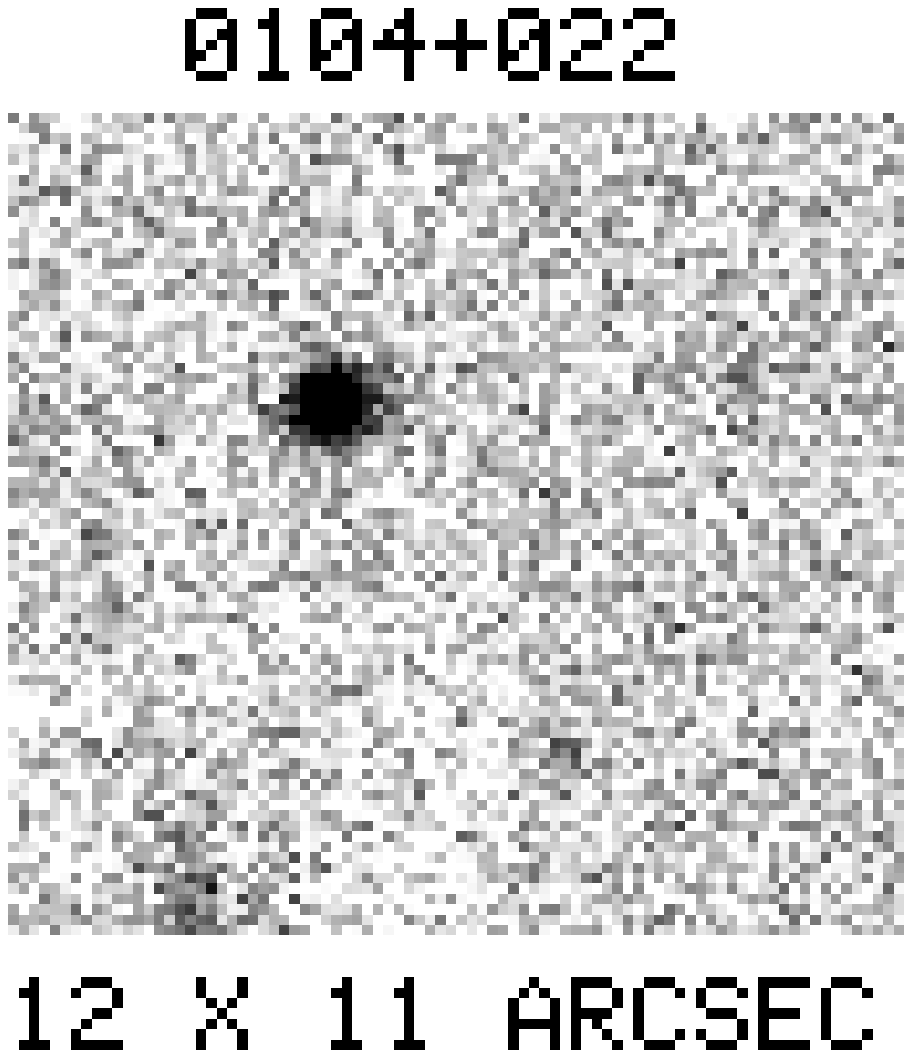]{Deep K-band image around QSO 0104+022. The QSO
is the bright object, which is extended almost E-W (the guide star lies to 
the south, and the lower left contains its diffraction spike). The QSO also
appears to have a curved filament to the NE. Faint objects
which may be companions lie at about equal distance to the W and SW.
\label{fig10}}

\end{document}